\documentclass[graybox]{svmult}

\usepackage{type1cm}
\usepackage{makeidx}
\usepackage{graphicx}
\usepackage{multicol}
\usepackage[bottom]{footmisc}
\usepackage{newtxtext}
\usepackage{booktabs}
\usepackage{array}
\usepackage[backend=biber, style=numeric, citestyle=numeric, maxbibnames=999, maxcitenames=999]{biblatex}
\usepackage[hidelinks]{hyperref}

\makeindex

\begin{document}

\title*{Qualitative Data Analysis in Software Engineering: Techniques and Teaching Insights}
\author{Christoph Treude}
\institute{Christoph Treude \at Singapore Management University, Singapore \email{ctreude@smu.edu.sg}}

\maketitle

\abstract{Software repositories are rich sources of qualitative artifacts, including source code comments, commit messages, issue descriptions, and documentation. These artifacts offer many interesting insights when analyzed through quantitative methods, as outlined in the chapter on mining software repositories. This chapter shifts the focus towards interpreting these artifacts using various qualitative data analysis techniques. We introduce qualitative coding as an iterative process, which is crucial not only for educational purposes but also to enhance the credibility and depth of research findings. Various coding methods are discussed along with the strategic design of a coding guide to ensure consistency and accuracy in data interpretation. The chapter also discusses quality assurance in qualitative data analysis, emphasizing principles such as credibility, transferability, dependability, and confirmability. These principles are vital to ensure that the findings are robust and can be generalized in different contexts. By sharing best practices and lessons learned, we aim to equip all readers with the tools necessary to conduct rigorous qualitative research in the field of software engineering.}

\section{Introduction}

In the area of software development, understanding and analyzing qualitative data from software repositories can produce many valuable insights. These repositories are rich with insightful artifacts, such as source code comments, commit messages, issue descriptions, and documentation~\cite{Kagdi2007}. Following the mining processes detailed in a previous chapter, this chapter focuses on how to interpret and analyze these artifacts using qualitative data analysis methods, with a special emphasis on teaching these methods.

Qualitative data analysis extends beyond mere data collection; it involves a thorough and iterative process of coding and interpretation~\cite{Seaman1999}. This chapter introduces qualitative coding not just as a technique, but as a systematic process integral to educational practice in software engineering. We explore various coding methods and emphasize the importance of designing a robust coding guide~\cite{Saldana2015}, tailored for researchers and educators.

We target an audience that includes students, researchers, software developers, and especially educators, providing them with the necessary tools and insights to effectively analyze qualitative data and teach these skills. Furthermore, the chapter stresses the importance of quality assurance in qualitative data analysis, highlighting key principles such as credibility, transferability, dependability, and confirmability~\cite{Lincoln1985} to ensure that the analysis is not only thorough, but also sound. We also address common challenges and pitfalls in qualitative data analysis, offering practical advice and best practices specifically tailored for educational settings~\cite{Creswell2016}. Lastly, the chapter discusses the nuances of teaching qualitative data analysis in software engineering, sharing lessons learned, and effective teaching practices.

\subsection{Target audience}

This chapter is particularly designed for educators in software engineering, providing them with a comprehensive guide to teaching qualitative data analysis. The reader will understand the significance of qualitative data in software engineering and will explore various types of software engineering artifact that can be analyzed using qualitative coding. The chapter offers a detailed description of the qualitative coding process and includes instructions on designing an effective coding guide. It emphasizes the importance of quality assurance in qualitative data analysis, providing checklists to ensure rigor and reliability. In addition, common mistakes and best practices are discussed to improve educators' ability to instruct effectively. Educators will not only gain insight into applying these methods, but also acquire practical tools to inspire and engage their students in adopting these techniques in their own projects.

\subsection{Learning Objectives}

Upon completing this chapter, readers should be able to:

\begin{itemize}
\item Identify and understand different qualitative artifacts commonly found in software engineering.
\item Apply systematic coding techniques to qualitative data to derive meaningful insights, suitable for classroom instruction.
\item Design and implement a coding guide tailored to both educational purposes and specific research needs.
\item Recognize and ensure the quality aspects of qualitative analysis, such as credibility and dependability, essential for teaching sound research methods.
\item Navigate common challenges in qualitative data analysis and employ best practices to overcome them.
\end{itemize}

\subsection{Teaching Strategies}

To effectively teach the content of this chapter, the following exemplary strategies are recommended for educators:

\begin{description}
\item[Interactive Workshops:] Facilitate interactive coding sessions in classrooms where students can practice coding real data extracted from software repositories. This hands-on approach helps cementing theoretical knowledge through practical application.
\item[Case Studies:] Use case studies from published papers to illustrate the application and results of qualitative coding. These examples can serve as a discussion base for understanding complex concepts and their implications in real world settings.
\item[Peer review:] Encourage students to peer review each other’s coding projects to foster a collaborative learning environment and critical thinking skills. This method also improves their ability to constructively critique and improve qualitative analyses.
\item[Role-Playing:] Implement role-playing exercises where students assume various roles involved in qualitative data analysis, such as data collector, coder, and quality assurer. This strategy helps students understand the collaborative and interdisciplinary nature of software engineering projects.
\end{description}

\section{Qualitative Artifacts in Software Repositories}

Software repositories are not just codebases; they are intricate ecosystems full of valuable information. Over the past decades, software analytics and mining software repositories have evolved into significant fields of research and practice~\cite{Hassan2008}. These activities involve sifting through large amounts of data, uncovering patterns, and extracting actionable insights to improve the software development process.

Although quantitative analysis has been the traditional focus, offering valuable insights, such as using metrics such as code complexity or line counts to predict defects~\cite{Bird2009}, there is growing recognition of the untapped potential in qualitative data. Qualitative artifacts, such as source code comments, commit messages, bug reports, developer conversations, interactions with language learning models (LLMs), and documentation, hold a wealth of information that can provide a deeper contextual understanding of the software development process. Each of these artifacts brings a different lens to the analysis, contributing to a multifaceted understanding of both technical and human factors in software projects. The following are a few concrete examples.

\textbf{Source Code Comments and Commit Messages:}
Source code comments and commit messages are often overlooked sources of information. Qualitative analysis here can uncover the narrative behind complex code changes~\cite{Buse2010}. For instance, a series of commit messages might reveal a developer's struggle with a particular bug, showcasing the iterative process of trial and error before the final solution. Comments in the code might also highlight areas where developers felt uncertain or where they anticipated future changes, indicating potential hotspots for future errors or refactoring needs. This analysis can guide project managers and team leaders in understanding team challenges, leading to targeted interventions or knowledge sharing sessions. The primary challenge in qualitatively analyzing these artifacts lies in their inconsistency and brevity. Developers may not consistently comment on their code or may write commit messages that are cryptic or too concise, making it difficult to derive meaningful insights without extensive contextual knowledge. For example, a cryptic commit message like ``big change'' appears frequently in repositories and often lacks a detailed explanation, complicating the task of tracing specific changes and their rationales. At the time of writing, a GitHub search for commit messages containing ``big change'', returned more than a million hits: \url{https://github.com/search?q=%22big+change%22&type=commits}

\textbf{Bug reports:}
Bug reports, when qualitatively analyzed, can offer much more than the identification of glitches~\cite{Bettenburg2008}. They can serve as a narrative of the user's journey, pinpointing not just technical failures but also misalignments between user expectations and software functionality. For example, a consistent theme in bug reports might not indicate a code defect but rather a feature that is counterintuitive to users, necessitating a redesign rather than a simple fix. Furthermore, the language used in these reports can indicate the severity and emotional impact of bugs on users, providing a measure of user satisfaction and urgency for fixes. The challenge in analyzing bug reports qualitatively is their sheer volume and variability. Each report may vary greatly in detail and quality, which requires researchers to sift through potentially hundreds of reports to find relevant data. As an illustrative point, searches for vague complaints like ``nothing works'' currently yield more than 100,000 results on GitHub (\url{https://github.com/search?q=%22nothing+works%22&type=issues}), each potentially varying greatly in context and severity, thus complicating the coding process.

\textbf{Developer Conversations:}
Developer conversations, found in forums, pull requests, or even chat logs, are rich in insight into team dynamics and decision-making processes~\cite{Storey2016, hata2022github}. Qualitative analysis of these conversations can highlight how decisions are made, whether there is a dominant voice that guides decision making, or whether there is a healthy debate before reaching a consensus. This analysis can also reveal how knowledge is shared within the team, potentially uncovering bottlenecks or knowledge silos that could impact the team's efficiency and effectiveness. The difficulty in analyzing developer conversations lies in the informal and unstructured nature of these interactions. They often contain a mix of technical jargon, slang, and shorthand, all of which can vary widely between different communities or even teams within the same community. To illustrate the challenges encountered when qualitatively analyzing developer conversations, many discussions on platforms such as Stack Overflow~\cite{treude2011programmers} remain unanswered, presenting a challenge in qualitative analysis as it may not be clear whether issues were resolved outside the forum or not, leaving the actual impact of such discussions ambiguous.

\textbf{Developer Interactions with LLMs:}
The advent of LLMs in software development creates new forms of data through developer interactions with these models~\cite{Xiao2024}. Qualitative analysis of these interactions can provide insight into the common challenges developers face, the kind of support they seek from AI, and the gaps in existing documentation or resources. For example, frequent queries about a specific feature or function can signal the need for better documentation or tutorials. Additionally, the tone and framing of queries can reflect user satisfaction or frustration, providing insights into the AI's utility and potential areas for improvement. Analyzing interactions with LLMs presents unique challenges, primarily due to the evolving nature of both the questions posed by developers and the responses generated by AI. This dynamism requires continuous updates to the analytical framework to ensure relevancy and accuracy in interpreting the data. For example, the DevGPT dataset~\cite{Xiao2024} which was curated to explore how software developers interact with ChatGPT, contains a total of almost 30,000 prompts / answers in almost 5,000 conversations, clearly showing that most conversations take many turns and require thorough analysis to understand the evolving needs of developers and AI responses.

To truly benefit from these qualitative artifacts, it is crucial to approach their analysis systematically. Proper qualitative data analysis can uncover the stories behind the data, providing insights that are not immediately apparent through quantitative methods alone. This requires a detailed understanding of the coding process, an appreciation of the context within which the software is developed, and a keen eye for the subtleties in the data. Successful applications of qualitative analysis in software engineering research illustrate how combining these different artifacts can lead to richer, more holistic insights into software development practices and challenges.

For example, a study by Fan et al., titled ``~`My GitHub Sponsors profile is live!' Investigating the Impact of Twitter/X Mentions on GitHub Sponsors'', explores the interplay between social media interactions and financial support mechanisms in open-source software~\cite{Fan2024}. By qualitatively analyzing tweets linking to GitHub Sponsors profiles, the study highlights how these social media endorsements can significantly impact sponsorship outcomes, providing a deeper understanding of the socioeconomic factors influencing developer support on platforms like GitHub. This integration of data from different sources such as GitHub and Twitter exemplifies the potential of qualitative methods to reveal complex dynamics in software development. We will refer to examples of the qualitative coding conducted in this study throughout the remainder of this chapter.

\section{The Basics of Qualitative Coding}

Qualitative coding is a fundamental process in qualitative data analysis, especially in the context of software engineering, where textual data is abundant~\cite{Stol2016}. It involves categorizing and interpreting textual data to glean insights and patterns. This process is not linear; it is iterative, often requiring researchers to review data and codes multiple times, refine, and interpret as understanding deepens~\cite{Saldana2015}. This section introduces the basics of qualitative coding, discussing various coding methods such as open, axial, and selective coding, each of which plays a crucial role in qualitative analysis~\cite{Charmaz2014}.

\textbf{Iterative process:}
Qualitative coding is a dynamic process that involves moving between data and codes, ensuring that the understanding of the data evolves and deepens with each review~\cite{Saldana2015}. Initially, the codes may be broad and somewhat vague, but as the researcher revisits the data, these codes become more refined and precise. This iterative cycle is essential to ensure that the final set of codes represents the depth and complexity of the data. It allows for the identification of new patterns or themes that may not have been apparent in the initial readings and for the modification or merging of existing codes as the researcher's understanding of the data matures.

As an example, in the study mentioned previously by Fan et al.~\cite{Fan2024}, the first step of qualitatively coding tweets related to open source sponsorship was unstructured, taking note of anything that appeared to be of interest in the data. In a subsequent phase, the initial question of ``what is the nature of these tweets?'' was refined into several subquestions, including an investigation of the relationship between the tweet author and users on GitHub, the timing of the tweet in relation to activities on GitHub, the nature of responses to the tweets, and the type of content.

\textbf{Open Coding:}
Open Coding is where the coding process begins and is all about immersion in the data~\cite{Charmaz2014}. Researchers approach data with an open mind, allowing the data to speak for themselves. This stage is exploratory and creative, involving a thorough line-by-line analysis of the text. The aim is to label the data with as many codes as necessary to describe all the nuances and facets of the data. These codes are often descriptive and can be phrases or words that encapsulate what the researcher believes to be significant in the text. It is a process of breaking down the data into manageable coded pieces which can then be further analyzed and categorized.

An important consideration in qualitative coding is whether to start with a preexisting taxonomy or framework (deductive approach) or to build the categories from the ground up based on the data (inductive or grounded approach)~\cite{Thomas2006}. Using a pre-existing taxonomy can provide a structured starting point, especially useful in fields where established frameworks exist. It can help to organize the data more systematically from the outset and can guide the analysis towards specific areas of interest. However, it is important to remain open to emerging themes that may not fit neatly into the preestablished categories. On the other hand, starting without a pre-defined taxonomy allows the data to dictate the categories (grounded theory approach). This approach can be more flexible and open to novel insights, but may require more iterative refinement to develop a coherent categorization structure. Both approaches have their merits, and the choice largely depends on the research objectives, the nature of the data, and the theoretical context of the study. It is essential to recognize the influence of the chosen approach on the coding process and to be vigilant about not forcing the data into preconceived categories, thereby potentially overlooking novel or unexpected insights. Moreover, researchers may sometimes find it beneficial to blend both approaches, using a preexisting taxonomy as a starting point while remaining open to emerging themes and categories, thereby harnessing the strengths of both deductive and inductive reasoning.

In the aforementioned study, open coding of the content of tweets related to open source sponsorship brought up a wide range of codes. For example, a tweet such as ``You can now feed my coffee addiction on Github Sponsors. So that I can continue debugging your systemd issues on NixOS'' could be coded with anything from a very specific ``GitHub user using sponsorship income to buy coffee'' to a much wider ``support for open-source contributions''. This initial broad coding helps identify key themes such as financial support, motivation for contributions, and community engagement within the open-source ecosystem. As the analysis progresses, these broad categories can be further refined and subdivided into more specific themes, providing a detailed map of the various factors that influence open source sponsorship. The final coding guide included categories such as ``generic advertisement'', ``advertisement with new functionality'', and ``donation appreciation'' to capture a wide variety of content.

\textbf{Axial Coding:}
Axial coding takes the process a step further by beginning to organize these codes into meaningful categories or themes~\cite{Charmaz2014}. It is about making connections between different codes identified during open coding and seeing how they relate to each other. This stage often involves creating visual representations, such as diagrams or flow charts, to help one see the relationships between codes. Axial coding helps transform a large set of disparate codes into a smaller set of categories or themes, each representing a key aspect of the data. It is about understanding the `axis' around which the codes revolve, hence the term `axial'. This stage is crucial for transitioning from descriptive to analytical coding, shifting the focus from what the data are to what they mean.

\textbf{Selective Coding:}
Selective coding is the final integration and refinement of categories and themes identified during axial coding~\cite{Charmaz2014}. Here, the researcher selects one or several `core' categories and systematically relates all other categories to these core ones. This stage is about building a coherent and cohesive narrative or theoretical framework that explains the phenomena represented by the data. It involves determining which categories are central to the narrative and how other categories and codes support, enhance, or relate to this central theme. Selective coding is about distilling the complexity of data into a clear, understandable, and coherent story or framework. It is the stage in which the researcher begins to understand the larger picture and the deeper meanings within the data.

\section{Designing a Coding Guide}

Although coding guides are valuable tools in many qualitative research contexts, their necessity and application can vary depending on the specific methodologies and goals of a study. In scenarios where consistency between different analysts is critical---such as in large-scale studies or when preparing for systematic reviews---a well-defined coding guide is indispensable. It helps ensure that all researchers interpret and code the data in a uniform manner, which is essential for the reliability and validity of the findings~\cite{MacQueen1998}. In addition, a coding guide serves as a detailed manual that provides clear guidelines and definitions, facilitating a more structured and comprehensible analysis process~\cite{Rabiee2004}.

However, in more exploratory or interpretative forms of qualitative research, such as phenomenology or certain forms of ethnography, the use of a rigid coding guide might be less appropriate. In these cases, the flexibility of the coding process allows researchers to adapt and evolve their analytical frameworks based on emerging data insights, which is crucial to a deep understanding of complex human experiences and behaviors~\cite{hoda2021socio}. Here, a coding guide might instead take the form of a more flexible set of principles or thematic directions rather than strict coding rules.

Regardless of the approach, a coding guide can also be an invaluable training resource. It provides structured guidance and examples for new researchers or team members, ensuring that they are well equipped to contribute effectively to the research. In addition, as a form of documentation, a coding guide records the analytical decisions made during the study, improving the transparency and reproducibility of the research~\cite{Miles2014}. Whether used as a strict procedural document or as a flexible analytical aid, the role of a coding guide should be tailored to the specific needs and objectives of the qualitative analysis being conducted.

Creating a coding guide starts with a thorough understanding of the data. This understanding is critical to identify potential codes and themes. The codes and categories are then clearly and concisely defined, ensuring that each code captures the essence of what it represents without being ambiguous.

The guide structure comes next, with codes organized into a framework that shows the relationships between themes and categories. But a coding guide is more than its structure; it is also about setting clear rules for code application. These rules help researchers navigate ambiguities and make decisions when data do not fit easily into predefined categories.

Examples are essential to a coding guide. They help coders understand how to apply each code practically, clarifying definitions and rules with real-world references~\cite{Miles2014}. However, a coding guide is not fixed; it is tested against a sample of data to refine codes, definitions, and rules as needed.

The final step in creating a coding guide is documenting the entire process. This document tells the story of how the guide was developed, highlighting the thought and rigor put into it, which adds to the research's credibility and transparency.

In software development, coding guides can vary widely, each tailored to specific data types. For example, a Code Review Comments Guide categorizes feedback during code reviews, providing insight into discussion quality and dynamics. A Commit Messages Guide can reveal patterns in how changes are communicated. A Bug Report Analysis Guide can identify common user issues and expectations, while a Developer Interaction Guide can analyze team communication and decision-making patterns.

Designing a coding guide requires precision, insight, and foresight. It is essential to ensure that the qualitative analysis is systematic, structured, and insightful. A well-designed coding guide can decode complex data within software repositories, turning intricate narratives into insights that drive software development forward.

The complete coding guide in the example study mentioned above is included in the publication~\cite{Fan2024} and included entries such as:

\begin{itemize}
\item Advertisement with new functionality: "This tweet explicitly advertises the author’s own GitHub Sponsors profile while mentioning new functionality of an open-source project."
\item Generic advertisement: "This tweet advertises the tweet author’s own GitHub Sponsors profile (use this code if the tweet does not fit the other advertisement categories)."
\item Sponsor template: "This tweet contains GitHub’s template for advertising one’s own GitHub Sponsors profile: 'My GitHub Sponsors profile is live! You can sponsor me to support my open source work' with no or minor changes."
\end{itemize}

These examples showcase the use of concrete examples in a coding guide, providing coding instructions (use this code if ...) that help coders apply each code practically, thereby clarifying definitions and rules with real-world references. This approach not only aids in training new researchers or team members, but also ensures that the coding process remains transparent and consistent across different datasets and research objectives.

\section{Quality in Qualitative Data Analysis}

In qualitative research, traditional quality metrics used in quantitative research, such as validity and reliability, assume a different complexion. The inherent subjectivity and depth of understanding sought in qualitative analysis require a different set of criteria to assess quality~\cite{Lincoln1985}. Validity, crucial in quantitative research, often focuses on the accuracy of measurements and the effectiveness of statistical inference. However, in qualitative research, where understanding and interpretation are central, the notion of validity does not quite capture the essence of what quality entails. Instead, qualitative researchers focus on the trustworthiness and rigor of their studies, which is where credibility, transferability, dependability, and confirmability come into play~\cite{Lincoln1985}.

\textbf{Credibility:}
Credibility refers to the confidence that can be placed in the truth of research findings~\cite{Morse2002}. In qualitative research, this is similar to the concept of internal validity in quantitative research, but with a focus on the richness, depth, and relevance of data and analysis. Achieving credibility involves prolonged engagement with the data, ensuring that the findings are deeply rooted in the data itself. Techniques such as member verification, where participants review and validate findings, and triangulation, where multiple data sources or methods are used to cross-verify the findings, improve the credibility of the research~\cite{Morse2002}.

\textbf{Transferability:}
Transferability refers to the extent to which the findings of qualitative research can be applied in other contexts or with other groups~\cite{Geertz1973}. Unlike the quantitative concept of external validity, transferability does not imply a broad generalizability. Instead, it acknowledges the uniqueness of each context and focuses on providing rich and detailed descriptions of the research context and the participants, allowing others to judge the applicability of the findings to their own contexts~\cite{Geertz1973}. A detailed description is a key strategy, providing a comprehensive account of the research setting and assumptions, allowing others to fully understand the context and assess the potential transferability of the findings.

\textbf{Dependability:}
Dependability is concerned with the consistency of research findings over time and is similar to the notion of reliability in quantitative research~\cite{Rodwell1998}. However, given the dynamic and evolving nature of qualitative research, dependability does not imply stability, but rather an understanding of how and why the findings might change. Ensuring dependability involves creating an audit trail, detailed documentation of the research process, decisions, and changes that occur during the study~\cite{Rodwell1998}. This allows for an external audit, where an independent examiner reviews the process and the findings, ensuring that the research was conducted rigorously and that the findings are well-substantiated.

\textbf{Confirmability:}
Confirmability is the qualitative counterpart to objectivity in quantitative research~\cite{Finlay2002}. It refers to the degree to which the findings are shaped by the respondents and not by researcher bias, motivation, or interest. This is achieved through reflexivity, where researchers continuously examine and account for their own biases, preferences, and influence throughout the research process~\cite{Finlay2002}. A reflexive journal, where researchers document their reflections, decisions, and challenges, can be a valuable tool for enhancing confirmability, ensuring that the findings accurately represent the data and not the researcher's pre-conceptions.

In addition to these criteria, it is sometimes beneficial to quantify the reliability of qualitative analysis by calculating inter-rater agreement. This involves having multiple researchers code the same data independently and then comparing how consistently they apply the codes and categories~\cite{Lincoln1985}. A high level of agreement among different raters can reinforce the dependability of the coding process, indicating that the interpretations are not solely dependent on the perspective of a single researcher. However, it is essential to balance this quantitative measure with the qualitative insights and depth that are central to qualitative research. The goal is not to force a quantitative framework onto qualitative data, but to use inter-rater agreement as one of several tools to enhance the trustworthiness and rigor of the analysis.

\section{Practical Approaches and Tools for Analysis}

In qualitative data analysis, the tools and methods used to examine and understand the data play an important role. However, it is important to note that the depth and quality of the research depend largely on the systematic and thoughtful approach of the researcher. This section discusses the key features of qualitative analysis tools, highlighting that simpler tools, such as Excel or pen-and-paper methods, can sometimes be most effective.

A good qualitative analysis tool should complement the researcher's work, offering features that align with the project's needs, and enhance the analysis process. Essential features include strong data organization and management. The tool should serve as a reliable data repository, supporting efficient categorization, coding, tagging, and retrieval. Visualization features are another important component, as they can reveal patterns and trends that might not be immediately apparent in the raw data. The ability to represent data in formats such as coding trees, frequency graphs, and relationship networks can provide quick and clear insights.

Collaboration is also a vital part of qualitative analysis in complex fields like software engineering. Tools that promote teamwork, allow shared projects, and support clear communication can significantly improve analysis. Features that enable tracking of changes, such as comment functions and audit trails, can provide valuable information and enhance the collective understanding of the data.

The tool's ability to handle various types of data, from textual bug reports to forum discussions and code comments, is also crucial. Software data often includes mixed text, comprising both natural language and code snippets. It is essential that the tool be versatile and capable of accurately representing this mixed-text format. Tools that cannot effectively manage and display the interplay between natural language and code may not be well suited for analyzing software data. Therefore, the selected tool should be capable of integrating and analyzing different data formats, including mixed text, to offer a comprehensive analysis platform.

However, it is essential to remember that the complexity of a tool does not necessarily equate to a more profound or accurate analysis. Tools should simplify and streamline the process, not complicate it or distract from the primary data. Sometimes, simpler tools can be the most effective. For example, Excel can be highly efficient for organizing, coding, and analyzing data due to its straightforward interface. Traditional methods such as the use of highlighters, sticky notes, and pen and paper can also be highly effective, offering a direct and engaging way to interact with data.

In the end, the true quality of qualitative analysis is not determined by the tools used but by the researcher's thorough and thoughtful approach. The ideal tool or method should integrate smoothly into the analytical process, enabling the researcher to deeply understand and interpret the data. The effectiveness of a tool is measured by its ability to become an integrated part of the researcher's cognitive and analytical processes, allowing a deeper and more nuanced understanding of the data.

Although simpler tools are often highly effective, there are several specialized software tools designed for more complex qualitative data analysis tasks. Here are a few notable examples:

\begin{itemize}
\item NVivo (\url{https://lumivero.com/products/nvivo/}): This powerful tool for qualitative and mixed-methods research helps organize and analyze non-numerical or unstructured data. It supports deeper data analysis and helps to reveal trends and patterns with features for coding, querying, and linking different data sets.
\item ATLAS.ti (\url{https://atlasti.com/}): Known for its robust capabilities for qualitative coding and analysis, this tool is suitable for a wide range of data types, including text, multimedia, and geographic or spatial data. It features a user-friendly interface and visual tools to map connections between themes.
\item MAXQDA (\url{https://www.maxqda.com/}): Ideal for software engineering projects that may involve diverse data sources, MAXQDA supports text, audio, and video data. It offers advanced coding tools, multimedia file management, and visualization options that aid in the structured analysis of qualitative data.
\item Dedoose (\url{https://www.dedoose.com/}): A web-based application providing features for mixed methods research, Dedoose excels in projects requiring both qualitative and quantitative analysis. Its key strengths include ease of use, affordability, and strong data visualization capabilities.
\item HyperRESEARCH (\url{https://www.researchware.com/products.html}): Known for its flexibility, this tool supports a variety of text and media data formats and offers a straightforward approach to coding and retrieving data. It also allows for the building of complex filters and reports.
\end{itemize}

The integration of Large Language Models (LLMs) into software engineering research presents significant opportunities and challenges~\cite{bano2024large}. LLMs, such as GPT-4 and ChatGPT, are becoming established in various disciplines, including qualitative software engineering research, where they offer the potential to enhance the research process. These models can optimize certain aspects of research, such as data coding and analysis, helping researchers handle large volumes of data more efficiently.

However, the adoption of LLMs in qualitative research also brings concerns about their ability to fully replace human roles, a notion debunked by empirical findings. The future of qualitative software engineering research is envisioned as a collaborative one, where LLMs and human researchers work together to advance the field. While LLMs show promise in aiding the qualitative research process, the critical and irreplaceable role of human researchers remains central. Humans are essential for ensuring ethical conduct, maintaining the validity and reliability of research findings, and managing the appropriate dissemination of results.

Furthermore, while LLMs can enhance the efficiency of data analysis, their limitations in capturing the nuanced understanding inherent to human researchers are notable. The ``human touch'' remains vital in interpreting and understanding qualitative data, a sentiment echoed in all anthropological and sociological disciplines. Furthermore, ethical considerations surrounding the use of LLMs, including issues of data privacy and intellectual property rights, require rigorous scrutiny to ensure that their integration into research practices adheres to the highest standards of ethical conduct.

\section{Common Pitfalls and Best Practices}

Educators who teach qualitative data analysis in software engineering should be particularly aware of potential pitfalls and adhere to best practices. This structured approach ensures that the analysis taught is not only insightful, but also methodologically sound~\cite{Maxwell2012}. In the following, we outline common pitfalls along with the corresponding best practices, emphasizing the need for a balanced and rigorous approach in the educational context.

\textbf{Confirmation Biases:}
One of the most common pitfalls is confirmation bias, where researchers may subconsciously favor data that align with their initial hypotheses, neglecting data that contradict them~\cite{Mays2000}. This bias can significantly skew the analysis. Educators should stress the importance of approaching data with an open mind, actively seeking disconfirming evidence, and engaging multiple researchers to provide diverse perspectives. This collective approach can offer a balanced interpretation that minimizes the influence of individual biases.

\textbf{Data Overload:}
Researchers often face the challenge of data overload, where the vast volume of data can lead to superficial analysis or overlooked nuances~\cite{Richards2005}. For educators, it is crucial to teach effective data management strategies. Organizing data into manageable segments, employing a robust coding framework, and focusing on depth rather than breadth in analysis can prevent being overwhelmed. Furthermore, selective and purposeful data sampling, focusing on data that are most relevant to the research questions, ensures clarity and focus in the analysis.

\textbf{Subjectivity:}
The subjective nature of qualitative analysis, while a strength, can also be a pitfall if it leads to overly subjective interpretations. Educators must guide students to improve objectivity and credibility by establishing clear and transparent processes. Documenting the decision-making process, providing well-defined codes with examples, and maintaining a reflexive journal to note assumptions and biases can improve the objectivity and traceability of the analysis.

\textbf{Inconsistency in Coding:}
Maintaining consistency in coding, particularly in collaborative projects, can be challenging. Inconsistencies in coding can cause confusion and compromise the integrity of the analysis~\cite{Krippendorff2004}. Best practice involves holding regular team discussions to align on the coding framework, employing intercoder reliability measures, and creating a detailed and comprehensive coding guide. These steps ensure that the coding remains consistent and reliable, regardless of the researcher.

\textbf{Over-Interpretation of Data:}
Researchers might fall into the trap of overinterpreting the data, reading more into it than is actually there. This can lead to conclusions that are not strongly supported by the data. Educators should emphasize the maintenance of a balance between creativity and discipline. Ensure that interpretations are directly linked to the data. Peer debriefing, where researchers discuss their interpretations with colleagues, can provide a check against overinterpretation, ensuring that conclusions are well-grounded in the data.

\textbf{Ignoring Negative Cases:}
Researchers can ignore or underestimate data that do not fit emerging patterns or themes, known as negative or deviant cases~\cite{Flyvbjerg2006}. This can lead to a one-sided understanding of the data. Educators should emphasize the importance of actively looking for and considering negative cases. These instances can provide valuable information and help refine and challenge emerging theories or patterns. Incorporating negative cases strengthens the analysis by providing a more comprehensive understanding of the data.

\textbf{Lack of Transparency:}
Without a clear exposition of the methods and processes used in data collection and analysis, research can lack transparency, making it difficult for others to understand, evaluate, or replicate the study~\cite{Lincoln1985}. Educators must emphasize the necessity of maintaining an audit trail. An audit trail includes detailed records of the data collection methods, the analysis process, the decision-making points, and any changes made throughout the investigation. It should also document the rationale behind methodological choices, coding decisions, and the evolution of the researcher's understanding. By preserving this level of detail, an audit trail not only enhances the transparency of the investigation, but also substantiates its credibility, allowing other researchers to follow the progression of the study and understand the context and reasoning behind the findings.

\textbf{Overemphasis on Tools Rather Than Concepts:}
Educators can fall into the trap of focusing too heavily on teaching specific qualitative analysis tools, neglecting the foundational concepts that underpin qualitative analysis. This can lead to students who are proficient in using tools, but may lack the ability to conduct analysis independently of specific software. Educators should balance tool instruction with robust discussions on theoretical frameworks and the principles of qualitative research to ensure that students understand the 'why' behind the 'how'.

\textbf{Lack of Real-World Examples:}
It is vital for educators to incorporate real-world examples to bridge theory with practice. Using simplified or hypothetical cases can fail to convey the complexity and nuances of actual qualitative analysis. Incorporating case studies from recent research, including pitfalls and successes, can provide students with a more realistic perspective on the challenges and dynamics of qualitative data analysis.

\textbf{Neglecting the Ethical Dimensions:}
Qualitative data often involve sensitive information, and ethical considerations are paramount. Educators may sometimes overlook the importance of discussing the ethical implications involved in data collection, analysis, and reporting. It is essential for educators to integrate ethics into the curriculum, ensuring that students understand how to ethically handle data, respect participant confidentiality, and report findings responsibly.

\textbf{Getting Overwhelmed by Rich Data:}
Qualitative data can be rich and complex, and researchers may feel overwhelmed by depth and volume, which could lead to burnout or analysis paralysis. Best practice is for educators to teach students to take a phased approach to the analysis, breaking the process into manageable stages. Regular breaks and reflections can help to maintain focus and perspective. Using visualizations and tools can also help to manage and organize data effectively.

In essence, while qualitative analysis presents unique challenges, being aware of these pitfalls and adhering to best practices can lead to insightful, credible, and valuable analysis. It is about striking the right balance between the richness of subjective insights and the rigor of systematic, methodical analysis, ensuring that the conclusions drawn from qualitative data are not only profound and nuanced, but also methodologically robust and reliable.

\section{Lessons Learned about Teaching Qualitative Data Analysis}

Teaching qualitative data analysis, particularly in the dynamic field of software engineering, offers a unique set of challenges and opportunities. A fundamental lesson is the importance of contextual learning. Students often understand the concepts of qualitative analysis more deeply when these concepts are framed within real-world scenarios and applications. Therefore, it is recommended to integrate case studies, use actual data from software projects, and refer to recent research within the curriculum. This method not only makes the learning experience more engaging, but also clearly demonstrates the practical relevance and application of qualitative analysis skills in real-world settings.

Encouraging critical thinking is crucial in qualitative analysis, a field characterized by its interpretative and subjective nature. Students should be nurtured to develop the ability to critically review data and to deeply reflect on their analytical choices and processes. As a best practice, educators should design courses that force students to justify their coding decisions, interpret data with care, and contemplate alternative perspectives. Students should be asked to question their preconceptions and recognize how their biases might color their analysis.

Understanding the iterative nature of qualitative data analysis is another key lesson. Unlike linear progression often portrayed in textbooks, qualitative analysis is a recursive process, involving continuous cycles of coding, categorization, and theme development. To impart this lesson effectively, educators should provide students with practical opportunities to participate in this iterative process. Assignments that require revisiting and refining analyses, supplemented with stages of constructive feedback, can demystify the iterative nature of qualitative analysis and cultivate students' skills in this area.

Although the intellectual process of analysis is crucial, familiarity with various qualitative data analysis tools can substantially increase the students' analytical efficiency and depth. Educators are encouraged to introduce students to a spectrum of tools, from basic software such as Excel to more sophisticated qualitative analysis programs. However, it is important that the focus on tools does not overshadow the foundational principles and practices of qualitative analysis.

The ability to collaborate effectively and communicate the findings clearly is important in qualitative analysis, which often involves teamwork. To foster these essential skills, educators are recommended to incorporate group projects and peer review sessions into their courses. Emphasis should be placed on the importance of transparency in the research process and the ability to articulate one's analytical approach and insights clearly and coherently.

Lastly, understanding and adhering to ethical standards is a critical component of qualitative research, particularly when human subjects are involved. Educators must ensure that students not only grasp the ethical dimensions of their research but also know how to practically apply these principles. This includes managing confidentiality, navigating the informed consent process, and handling data responsibly. Integrating research ethics modules into the curriculum can be beneficial. These modules could include real-world examples and case studies that illustrate ethical dilemmas that researchers often face. In addition, offering practical exercises, such as drafting consent forms or developing a data management plan, can provide students with hands-on experience in addressing these ethical issues.

\section{Quality Checklist}

\begin{table}[!ht]
\caption{Qualitative Analysis Checklist}
\centering
\begin{tabular}{@{}lp{7.4cm}@{}}
\toprule
\textbf{Stage} & \textbf{Checklist Item} \\ \midrule
\textbf{Data Familiarization} & 
\parbox{7.4cm}{
    \noindent\hangindent=1em\hangafter=1 Have you thoroughly immersed yourself in the data? \par
    \noindent\hangindent=1em\hangafter=1 Have you continuously reviewed the data to deepen your understanding?} \\ \midrule
\textbf{Open Coding} & 
\parbox{7.4cm}{
    \noindent\hangindent=1em\hangafter=1 Have you started coding without pre-conceived notions, allowing the data to inform your codes? \par
    \noindent\hangindent=1em\hangafter=1 Are your codes descriptive and deeply rooted in the data?} \\ \midrule
\textbf{Axial Coding} & 
\parbox{7.4cm}{
    \noindent\hangindent=1em\hangafter=1 Have you examined the relationships between codes and organized them into coherent categories? \par
    \noindent\hangindent=1em\hangafter=1 Are you developing a narrative or framework that connects these categories meaningfully?} \\ \midrule
\textbf{Selective Coding} & 
\parbox{7.4cm}{
    \noindent\hangindent=1em\hangafter=1 Have you pinpointed core categories and related all other categories to these? \par
    \noindent\hangindent=1em\hangafter=1 Is there a clear, overarching narrative or theory emerging from your analysis?} \\ \midrule
\textbf{Iterative Refinement} & 
\parbox{7.4cm}{
    \noindent\hangindent=1em\hangafter=1 Are you revisiting your codes and categories, refining, and redefining them as understanding matures? \par
    \noindent\hangindent=1em\hangafter=1 Is the iterative process contributing to a deeper, more nuanced understanding of the data?} \\ \midrule
\textbf{Collaboration and Peer Review} & 
\parbox{7.4cm}{
    \noindent\hangindent=1em\hangafter=1 Have you engaged with peers or mentors to discuss your codes and findings? \par
    \noindent\hangindent=1em\hangafter=1 Are you open and receptive to feedback, willing to reconsider aspects of your analysis based on this input?} \\ \midrule
\textbf{Reflection and Reflexivity} & 
\parbox{7.4cm}{
    \noindent\hangindent=1em\hangafter=1 Have you maintained a record of your analytical decisions, reflections, and challenges faced? \par
    \noindent\hangindent=1em\hangafter=1 Are you critically examining your own biases and their influence on your interpretation?} \\ \midrule
\textbf{Ethical Consideration} & 
\parbox{7.4cm}{
    \noindent\hangindent=1em\hangafter=1 Have you ensured the confidentiality and ethical treatment of your data? \par
    \noindent\hangindent=1em\hangafter=1 Are you adhering to ethical standards and guidelines in your analysis and reporting?} \\ \midrule
\textbf{Transparency in Documentation} & 
\parbox{7.4cm}{
    \noindent\hangindent=1em\hangafter=1 Have you kept a clear and detailed record of your methodology and findings? \par
    \noindent\hangindent=1em\hangafter=1 Is your documentation complete enough to allow others to understand, evaluate, or replicate your study?} \\ \midrule
\textbf{Communication Findings} & 
\parbox{7.4cm}{
    \noindent\hangindent=1em\hangafter=1 Are you presenting your findings clearly and logically, articulating the developed narrative? \par
    \noindent\hangindent=1em\hangafter=1 Have you provided ample evidence, like direct quotes or clear examples, to support your findings?} \\
\bottomrule
\end{tabular}
\label{tab:qual_analysis}
\end{table}

Summarizing the principles of qualitative data analysis in software engineering, we recognize the intricacies and depth involved in this analytical process. It requires a systematic and thoughtful approach to ensure that the analysis is thorough and methodologically sound. To support this process, we introduce the Quality Checklist shown in Table~\ref{tab:qual_analysis}. This checklist consists of key questions designed to guide students, researchers, and practitioners in maintaining rigor and attention to detail throughout each phase of their qualitative analysis.

\section{Conclusion}

Qualitative data analysis in software engineering entails a thorough examination and interpretation of data, transforming these insights into actionable information that can refine the software development process. This chapter has explored the rich variety of qualitative data present in software repositories, such as source code comments and developer interactions, each offering unique insights into the complexities of software projects.

Throughout this discussion, we have explored qualitative coding processes, including open, axial, and selective coding, and emphasized the critical importance of revisiting and reflecting on the data to deepen understanding and refine analysis. The iterative nature of this analysis is essential for developing a comprehensive understanding of the underlying patterns and themes.

A well-crafted coding guide has been highlighted as more than just a tool: It serves as a crucial instructional resource that helps researchers, particularly educators, navigate the complexities of qualitative data. This guide not only supports consistent and accurate coding practices, but also serves as an invaluable training tool for new researchers and students. By outlining clear guidelines and examples, it ensures that the analysis remains structured and transparent.

We have underscored the foundational qualities of qualitative research: credibility, transferability, dependability, and confirmability, stressing the importance of conducting studies that are both trustworthy and thorough. The discussion of practical tools and methods has noted that sometimes simpler approaches, such as using Excel or traditional pen-and-paper techniques, can be highly effective, especially in educational settings where resources may be limited.

Furthermore, this chapter has addressed common pitfalls and best practices in qualitative analysis, advocating for a structured and balanced approach that can yield insightful and reliable results. Critical insights on teaching qualitative data analysis have been shared, highlighting the need to provide real-world context, fostering critical thinking, and emphasizing the iterative nature of analysis. These elements are crucial in equipping students not only with technical skills, but also with a deeper understanding of the research process.

In conclusion, qualitative data analysis in software engineering is not merely data interpretation; it is about crafting a narrative that is methodologically sound, contextually rich, and deeply insightful. For educators, this field offers dynamic opportunities to engage students in a reflective and structured inquiry into the nuances of software development, propelled by insightful qualitative analysis. This chapter aims to serve as a foundational resource for educators striving to instill these critical skills in the next generation of software engineers, thereby enhancing both their understanding and their impact on the field.

\section{Further Readings}

\begin{itemize}
    \item \fullcite{Adolph2011}~\cite{Adolph2011}
    \item \fullcite{Baltes2022}~\cite{Baltes2022}
    \item \fullcite{Bryman2012}~\cite{Bryman2012}
    \item \fullcite{DeFranco2017}~\cite{DeFranco2017}
    \item \fullcite{Denzin2011}~\cite{Denzin2011}
    \item \fullcite{Dittrich2007}~\cite{Dittrich2007}
    \item \fullcite{Dyba2005}~\cite{Dyba2005}
    \item \fullcite{Easterbrook2008}~\cite{Easterbrook2008}
    \item \fullcite{Flick2018}~\cite{Flick2018}
    \item \fullcite{Glaser1967}~\cite{Glaser1967}
    \item \fullcite{Hammersley2007}~\cite{Hammersley2007}
    \item \fullcite{Hoda2010}~\cite{Hoda2010}
    \item \fullcite{Ko2010}~\cite{Ko2010}
    \item \fullcite{Lenberg2023}~\cite{Lenberg2023}
    \item \fullcite{Lenberg2017}~\cite{Lenberg2017}
    \item \fullcite{Ralph2018}~\cite{Ralph2018}
    \item \fullcite{ralph2020empirical}~\cite{ralph2020empirical}
    \item \fullcite{Shaw2002}~\cite{Shaw2002}
    \item \fullcite{Sjoberg2007}~\cite{Sjoberg2007}
    \item \fullcite{Stol2018}~\cite{Stol2018}
    \item \fullcite{Strauss1998}~\cite{Strauss1998}
    \item \fullcite{Wohlin2015}~\cite{Wohlin2015}
    \item \fullcite{Wohlin2006}~\cite{Wohlin2006}
\end{itemize}

\printbibliography

\end{document}